\documentclass[sigconf,screen,10pt,balance=false]{acmart}

\usepackage[symbol]{footmisc}
\usepackage{setspace}
\usepackage{multirow}
\usepackage{tcolorbox}
\tcbuselibrary{breakable}
\usepackage{balance}
\usepackage{amsmath}
\usepackage{centernot}
\usepackage{alltt}
\usepackage{listings}
\usepackage{breakurl}
\usepackage{fancyvrb}
\usepackage{relsize}
\usepackage{soul}
\usepackage{array}
\usepackage{booktabs}
\usepackage{enumitem}
\usepackage{currfile}
\usepackage{tikz}
\usepackage{pbox}
\usepackage{xstring}
\usepackage{xspace}
\usepackage{epstopdf}
\usepackage{graphicx}
\usepackage{float}
\usepackage{makecell}
\usepackage{url}
\usepackage{algorithm}
\usepackage{algpseudocode}
\usepackage{tabularx}
\usepackage{multirow}
\usepackage{makecell}
\usepackage{subcaption} 
\usepackage{xcolor}
\usepackage{algorithm}
\usepackage{algpseudocode}
\usepackage{graphicx}
\acmConference[CAIS '26 SAO Workshop]{CAIS '26 SAO Workshop}{May 26, 2026}{San Jose, CA}

\renewcommand\footnotetextcopyrightpermission[1]{}
\setcopyright{none}

\AtBeginDocument{%
  \providecommand\BibTeX{{%
    \normalfont B\kern-0.5em{\scshape i\kern-0.25em b}\kern-0.8em\TeX}}}

\algnewcommand\algorithmicforeach{\textbf{for each}}
\algdef{S}[FOR]{ForEach}[1]{\algorithmicforeach\ #1\ \algorithmicdo}
\makeatletter
\newcommand{\mybox}[1]{%
  \setbox0=\hbox{#1}%
  \setlength{\@tempdima}{\dimexpr\wd0+4pt}%
  \begin{tcolorbox}[colback={blue!50},boxrule=0.1pt,arc=2pt,
      left=1pt,right=0pt,top=1pt,bottom=1pt,boxsep=0pt,width=\@tempdima]
    #1
  \end{tcolorbox}
}
\makeatother

\makeatletter
\def\url@leostyle{%
  \@ifundefined{selectfont}{\def\UrlFont{\sf}}{\def\UrlFont{\small\ttfamily}}}
\makeatother
\floatstyle{boxed}
\newfloat{aside}{bth!}{lop}
\floatname{aside}{Aside}

 \usetikzlibrary{calc}
\makeatletter
\def\mfontsize{\f@size}

\makeatother

\definecolor{light-green}{RGB}{64, 219, 64}

\makeatletter
\newcommand{\sqbox}{%
    \collectbox{%
        \@tempdima=\dimexpr\width-\totalheight\relax
        \ifdim\@tempdima<\z@
            \fbox{\hbox{\hspace{-.5\@tempdima}\BOXCONTENT\hspace{-.5\@tempdima}}}%
        \else
            \ht\collectedbox=\dimexpr\ht\collectedbox+.5\@tempdima\relax
            \dp\collectedbox=\dimexpr\dp\collectedbox+.5\@tempdima\relax
            \fbox{\BOXCONTENT}%
        \fi
    }%
}
\makeatother

\begin{document}
%\pagenumbering{gobble}

% Haryadi .. parskip, and par-indent
%\setlength\parindent{0pt}
%\setlength\parskip{5pt}

\newcommand{\beforesec}{\vspace{-.4cm}}
\newcommand{\aftersec}{\vspace{-.3cm}}

\newcommand{\beforesect}{\vspace{-.1cm}}
\newcommand{\aftersect}{\vspace{-.3cm}}

\newcommand{\beforesub}{\vspace{-.3cm}}
\newcommand{\aftersub}{\vspace{-.25cm}}

\newcommand{\beforesubsub}{\vspace{-.2cm}}
\newcommand{\aftersubsub}{\vspace{-.2cm}}

 \newcommand{\zsection}[1]{\section{#1}}
 \newcommand{\zsubsection}[1]{\subsection{#1}}
 \newcommand{\zsubsubsection}[1]{\subsubsection{#1}}

\newcommand{\smush}{0.25in}

\newcommand{\figWidthOne}{3.05in} 
\newcommand{\figWidthHalf}{1.45in} 
\newcommand{\figWidthTwo}{3.05in} 
\newcommand{\figWidthTwop}{1.6in} 
\newcommand{\figWidthThree}{2.2in} 
\newcommand{\figWidthSix}{1.1in} 
\newcommand{\figWidthFour}{1.7in} 
\newcommand{\figHeight}{2.0in}
\newcommand{\captionText}[2]{\textbf{#1} \textit{{#2}}}

\newcommand{\zbullet}{\hspace{-0.1cm}$\bullet$}

\newcommand{\eg}{e.g.}
\newcommand{\ie}{i.e.}
\newcommand{\etal}{et al.}
\newcommand{\apriori}{\textit{a priori}}

%-------------------------------------------------------------------
% Haryadi -- new command
\newcommand{\msub}[1]{\vspace{1pt}\noindent{\bf #1}}

\newcommand{\ts}[1]{{\tt{\small#1}}}
\newcommand{\tsb}[1]{{\tt{\small{\bf#1}}}}
\newcommand{\tse}[1]{{\tt{\small{\em#1}}}}
\newcommand{\tss}[1]{{\tt{\footnotesize#1}}}
\newcommand{\exc}{$^{\ddag}$}        % except
\newcommand{\EIO}{\ts{EIO}}
\newcommand{\ENOSPC}{\ts{ENOSPC}}
\newcommand{\EDQUOT}{\ts{EDQUOT}}

%-------------------------------------------------------------------
% For fault injection table
\newcommand{\ip}{bad}
\newcommand{\nullp}{$\emptyset$}

\newcommand{\oops}{o}
\newcommand{\dead}{$\times$}
\newcommand{\alive}{$\surd$}
\newcommand{\unuse}{$\times$}
\newcommand{\use}{$\surd$}
\newcommand{\ic}{$\times$}
\newcommand{\con}{$\surd$}
\newcommand{\gpf}{G}
\newcommand{\npe}{null-pointer}
\newcommand{\usebuta}{$\surd^a$} % unmountable
\newcommand{\hwdetect}{d}

%% TABLE 2
\newcommand{\lateoops}{o$^b$}  % oops happens, but late
\newcommand{\lategpf}{G$^b$}   % gpf, but late
\newcommand{\iop}{i}           % invalid opcode
\newcommand{\detects}{d}       % assertion
\newcommand{\silentret}{s}     % app fails silently
\newcommand{\errorret}{e}      % app returns an error
\newcommand{\appworks}{$\surd$}     % app works!
\newcommand{\usebutar}{$\surd^{ar}$} % read-only and unmountable

\newcommand{\tbls}{\hspace{0.025in}}
\newcommand{\tblss}{\hspace{0.015in}}
\newcommand{\shrinkless}{\vspace{-0.01cm}}

%-------------------------------------------------------------------

%-------------------------------------------------------------------

% axes         
\newcommand{\x}{{\em x}}
\newcommand{\y}{{\em y}}
\newcommand{\xaxis}{x-axis}
\newcommand{\yaxis}{y-axis}

\newcommand{\KB}{~KB}
\newcommand{\KBs}{~KB/s}
\newcommand{\Kbs}{~Kbit/s}
\newcommand{\mbs}{~Mbit/s}
\newcommand{\MB}{~MB}
\newcommand{\GB}{~GB}
\newcommand{\MBs}{~MB/s}
\newcommand{\mus}{\mbox{$\mu s$}}
\newcommand{\ms}{\mbox{$ms$}}

\newcommand{\unix}{{\sc Unix}}

\newcommand{\bquote}{\vspace{-0.25cm} \begin{quote}}
\newcommand{\equote}{\end{quote}\vspace{-0.05cm} }

\newcommand{\zquote}[2]{\begin{quote}
#1 --
{\em``#2'' }
%{\bf -- #1 }
\end{quote}}

\newcommand{\XXX}[1]{{\small {\bf (XXX: #1)}}}

\newcommand{\XXXX}{{\bf XXX}}
\newcommand{\xx}{{\bf XXX}}

% normal
\newcommand{\beforecaption}{\begin{spacing}{1.2}}
\newcommand{\aftercaption}{\end{spacing}}
\newcommand{\mycaption}[3]{{\beforecaption\caption{\label{#1}{\bf \small #2. } {\em #3}}\aftercaption}}
\newcommand{\rcaption}[3]{{\beforecaption\caption{\label{#1}{\bf \small #2 } {\em #3}}\aftercaption}}

\newcommand{\sref}[1]{\S\ref{#1}}

\newcommand{\xxx}[1]{  \underline{ {\small {\bf (XXX: #1)}}}}

\newenvironment{packeditemize}{
\begin{itemize}[leftmargin=*]
  \setlength{\itemsep}{1pt}
  \setlength{\parskip}{0pt}
  \setlength{\parsep}{0pt}
}{\end{itemize}}

\newcommand{\llname}{{LazyLog}}
\newcommand{\llnamebig}{{LazyLog}}
\newcommand{\llnamesmall}{{LazyLog}}

\newcommand{\rulename}{order-when-externalized}
\newcommand{\rulenamecaps}{Order-when-externalized}

\newcommand{\noext}{nilext}
\newcommand{\noextcap}{Nilext}

\newcommand{\recmsg}{\mbox{\textsc{\fontsize{9.2}{5}\selectfont Recovery}}}
\newcommand{\recresmsg}{\mbox{\textsc{\fontsize{9.2}{5}\selectfont RecoveryResponse}}}
\newcommand{\svcmsg}{\mbox{\textsc{\fontsize{9.2}{5}\selectfont StartViewChange}}}
\newcommand{\dvcmsg}{\mbox{\textsc{\fontsize{9.2}{5}\selectfont DoViewChange}}}
\newcommand{\svmsg}{\mbox{\textsc{\fontsize{9.2}{5}\selectfont StartView}}}

\newcommand{\fwrong}{$\times$}
\newcommand{\smalltt}[1]{\texttt{\fontsize{8.7}{5}\selectfont #1}}
\newcommand{\smallit}[1]{\textit{\scriptsize #1}}
\newcommand{\verysmall}[1]{\scriptsize #1}
\newcommand{\verysmalltt}[1]{\texttt{\scriptsize #1}}
\newcommand*\rot{\rotatebox{90}}
\newcommand{\alps}{\mbox{\textsc{\fontsize{12}{5}\selectfont alps}}}
\newcommand{\cmap}{\mbox{\textsc{\fontsize{9.2}{5}\selectfont Masc}}}

% ctrl local storage layer
\newcommand{\ctrlstore}{\mbox{\textsc{\fontsize{9.2}{5}\selectfont Clstore}}}
\newcommand{\ctrlstoresmall}{\mbox{\textsc{\fontsize{7.2}{5}\selectfont Clstore}}}

\newcommand{\ctrlnamesmall}{\mbox{\textsc{\fontsize{7.2}{5}\selectfont Ctrl}}}
\newcommand{\ctrlnaive}{\mbox{\textsc{\fontsize{9.2}{5}\selectfont Ctrl-Naive}}} 
\newcommand{\ctrlnaivesmall}{\mbox{\textsc{\fontsize{7.2}{5}\selectfont Ctrl-Naive}}} 

\newcommand{\dfs}{{dfs}}
\newcommand{\dftname}{\mbox{\textsc{\fontsize{9.2}{5}\selectfont Dft}}} 
\newcommand{\dftnamesmall}{\mbox{\textsc{\fontsize{7.2}{5}\selectfont Dft}}} 

\newcommand{\sysbasic}{{Erwin-{\fontsize{7.2}{5}\selectfont{$\blacksquare$}}}}
\newcommand{\sysst}{{Erwin-st}}

\newcommand{\writeSC}{\smalltt{write()}}
\newcommand{\fsyncSC}{\smalltt{fsync()}}
\newcommand{\msyncSC}{\smalltt{msync()}}
\newcommand{\fdatasyncSC}{\smalltt{x fdatasync()}}
\newcommand{\linkSC}{\smalltt{link()}}
\newcommand{\mkdirSC}{\smalltt{mkdir()}}
\newcommand{\fempty}{$\phi$}
\newcommand{\fexists}{$\surd$}
\newcommand{\creatSC}{{\smalltt{creat()}}}
\newcommand{\unlinkSC}{{\smalltt{unlink()}}}
\newcommand{\renameSC}{{\smalltt{rename()}}}
\newcommand\floor[1]{\lfloor#1\rfloor}
\newcommand\ceil[1]{\lceil#1\rceil}
\newcommand{\totbugs}{60}
\newcommand{\totapps}{11}
\newcommand{\totappsw}{eleven}
\newcommand*{\combination}[2]{{}^{#1}C_{#2}}

\newcommand{\used}{$\surd$}
\newcommand{\usedpar}{$P$}
\newcommand{\useddollar}{$\surd$\textsuperscript{\$}}
\newcommand{\usedadler}{$\surd$\textsuperscript{a}}
\newcommand{\usedparstar}{$P$\textsuperscript{$*$}}
\newcommand{\notused}{}
\newcommand{\numapps}{eight}

\if 0 % colored version
\newcommand{\yes}{$\surd$}
\newcommand{\yesi}{\colorbox{gray!30}{$\surd$\textsubscript{$i$}}}
\newcommand{\nolow}{\colorbox{gray!30}{$\times$\textsubscript{$l$}}}
\newcommand{\no}{\colorbox{gray!85}{$\times$}}
\newcommand{\complower}{$L$}
\newcommand{\compmoder}{$M$}
\newcommand{\comphigher}{\colorbox{gray!85}{$H$}}
\newcommand{\na}{na}
\fi

\newcommand{\yes}{$\surd$}
\newcommand{\yesi}{$\surd$\textsubscript{$i$}}
\newcommand{\nolow}{$\times$\textsubscript{$l$}}
\newcommand{\nomod}{$\times$\textsubscript{$m$}}
\newcommand{\no}{$\times$}
\newcommand{\yeslow}{$\surd$\textsuperscript{$l$}}
\newcommand{\complower}{$L$}
\newcommand{\compmoder}{$M$}
\newcommand{\comphigher}{$H$}
\newcommand{\na}{na}

\newcommand*{\termindex}[2]{$\langle$\textit{epoch}:{#1}, \textit{index}:{#2}$\rangle$}
\newcommand*{\termindexnovar}{$\langle$\textit{epoch}, \textit{index}$\rangle$}
\newcommand*{\snapid}{$\langle$\textit{snap-index}, \textit{chunk}\#$\rangle$}
\newcommand*{\rafttermindexnovar}{$\langle$\textit{term}, \textit{index}$\rangle$}
\newcommand{\quotes}[1]{``#1''}
\newcommand{\parnew}[1]{{#1}}

\newcommand{\xh}[1]{{{\color{blue} [xuhao: #1]}}}
\newcommand{\ag}[1]{{{\color{red} [Aishwarya: #1]}}}
\newcommand{\sgnotes}[1]{{\color{blue} #1}}

\newcommand{\sysname}{\emph{Sophrosyne}}
\newcommand{\grn}[1]{\textcolor{green!60!black}{\textbf{#1}}}
\newcommand{\darkgrn}[1]{\textcolor{green!30!black}{\textbf{#1}}}
\newcommand{\hybrid}{\ensuremath{\mathbf{HI}_{\mathbf{col}}}}
\newcommand{\blfootnote}[1]{%
  \begingroup
  \renewcommand\thefootnote{}\footnote{#1}%
  \addtocounter{footnote}{-1}%
  \endgroup
}
\newcommand{\textttwrap}[1]{%
  {\ttfamily\hyphenchar\font=`\-\relax #1}%
}
\begin{abstract}
Text2SQL agents powered by LLMs translate natural language intent into SQL by \emph{exploring} the data system through tool calls before formulating the query. However, to ensure secure and scoped access, data systems construct environments with explicit API surfaces. We study and categorize these APIs exposed today as either \emph{coarse-grained} or \emph{fine-grained} and posit that choosing between them presents a fundamental tradeoff between cost-efficient exploration and accurate SQL generation. Most data systems expose fine-grained APIs, but this inadvertently disadvantages agents: they \emph{over-explore}, incorporating irrelevant schema elements into their query formulation and produce \emph{inaccurate} results. We argue that curbing over-exploration is key to the effective use of these API surfaces, and propose \sysname{}, a data system environment that augments API responses with \emph{directives} that guide the agent's exploration process. Initial results show that directives reduce over-exploration by 4.6$\times$ and boost accuracy by up to 12.4\% ($\sim$4\ percentage points).
\end{abstract}



\date{}

\title[Sophrosyne: Agentic Exploration of Relational Data Systems Needs Moderation]{\LARGE Sophrosyne: Agentic Exploration of Relational Data Systems Needs Moderation}
\author{\Large Madhav Jivrajani}
\affiliation{
    \institution{\emph{University of Illinois Urbana-Champaign}}
    \city{}
    \country{}
}

\author{\Large Ramnatthan Alagappan}
\affiliation{
    \institution{\emph{University of Illinois Urbana-Champaign}}
    \city{}
    \country{}
}

\author{\Large Aishwarya Ganesan}
\affiliation{
    \institution{\emph{University of Illinois Urbana-Champaign}}
    \city{}
    \country{}
}

\renewcommand{\shortauthors}{M. Jivrajani, R. Alagappan, A. Ganesan}

\maketitle
\thispagestyle{plain}
\blfootnote{\sysname{} is the ancient Greek virtue of prudence and moderation}
\vspace*{-8mm}
\section{Introduction}
\label{sec-intro}

With advanced tool calling abilities \cite{schick2023toolformerlanguagemodelsteach}, AI agents are emerging as the primary users of data systems \cite{liu2025supportingaioverlordsredesigning,zaharia2025bridging}, and text2SQL agents in particular offer a promising approach to translating natural language queries to SQL. Agents begin by \emph{exploring} \cite{liu2025supportingaioverlordsredesigning} the data system to gather relevant tables, columns, and \texttt{JOIN} relationships prior to formulating the final query.

\begin{figure}
    \centering
    \includegraphics[width=0.75\linewidth]{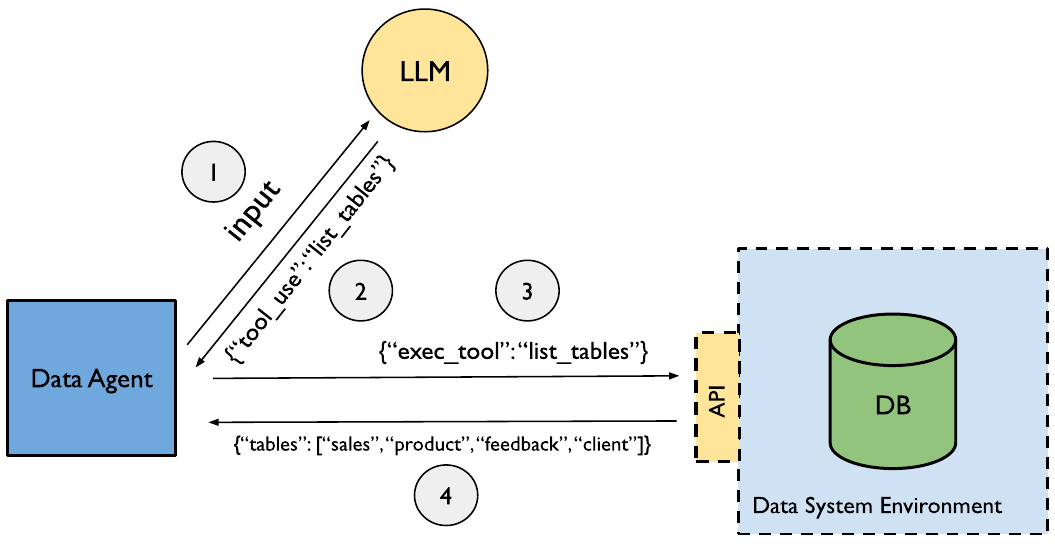}
    \caption{Agents and Data System Environments}
    \label{fig:agent_loop}
    \Description[Data system environment agentic loop]{Tool calling against a data system environment.}
\end{figure}

While exposing data systems to agents is powerful, securing and scoping access to them is crucial \cite{mcpcli, mcpauth}. As a result, data systems construct environments with explicit API surfaces and authentication \cite{google_spanner_mcp_2026, planetscale_mcp_2026, aws_aurora_dsql_mcp_2025, aws_mysql_mcp_2025, aws_postgres_mcp_2025, aws_redshift_mcp_2025, google_cloud_mcp_2025_bigquery, azure_dab_mcp_dml_2025, supabase_mcp_2025, neon_mcp_2026, motherduck_mcp_2025, turso_mcp_2026} using protocols such as MCP \cite{anthropic2024mcp}. Agents execute tool calls by invoking APIs exposed to it as shown in Figure \ref{fig:agent_loop}. Since this is becoming the ubiquitous way data systems are exposed to agents, we ask the question: \emph{how does the exposed API surface effect the agent's ability to perform text2SQL tasks?} We conduct the first study of API surfaces exposed by different data systems and find that they fall into two broad categories and differ primarily in how they enable exploration: \emph{coarse-grained} and \emph{fine-grained} APIs. \emph{Coarse-grained} APIs allow agents to retrieve  the entire schema up-front for the LLM to process in its entirety and determine relevant parts for query formulation. \emph{Fine-grained} APIs enable progressive exploration, starting with just table names and allow the agent to selectively inspect individual table schemas.

\begin{figure}[t]
    \centering
    % --- Vendor Overview (top) + Inefficiency Figure (bottom) ---
    \begin{minipage}[t]{\linewidth}
        \centering
        \captionof{table}{Categorization of API Surfaces}
        \label{tab:vendor-overview}
        \small
        \renewcommand{\arraystretch}{1.05}
        \scalebox{0.82}{
        \begin{tabularx}{\linewidth}{>{\bfseries}l|X}
            API Surface & \textbf{Data System} \\
            \hline
            \footnotesize Coarse-grained & Google Spanner \cite{google_spanner_mcp_2026}, PlanetScale \cite{planetscale_mcp_2026} \\
            \hline
            \footnotesize Fine-grained & Amazon DSQL \cite{aws_aurora_dsql_mcp_2025}, Amazon MySQL \cite{aws_mysql_mcp_2025}, Amazon Postgres \cite{aws_postgres_mcp_2025}, Amazon Redshift \cite{aws_redshift_mcp_2025}, GCP BigQuery \cite{google_cloud_mcp_2025_bigquery}, Azure SQL \cite{azure_dab_mcp_dml_2025}, Supabase \cite{supabase_mcp_2025}, Neon \cite{neon_mcp_2026}, MotherDuck DuckDB \cite{motherduck_mcp_2025}, Turso \cite{turso_mcp_2026} \\
        \end{tabularx}}

        \vspace{1em}

        \includegraphics[width=\linewidth]{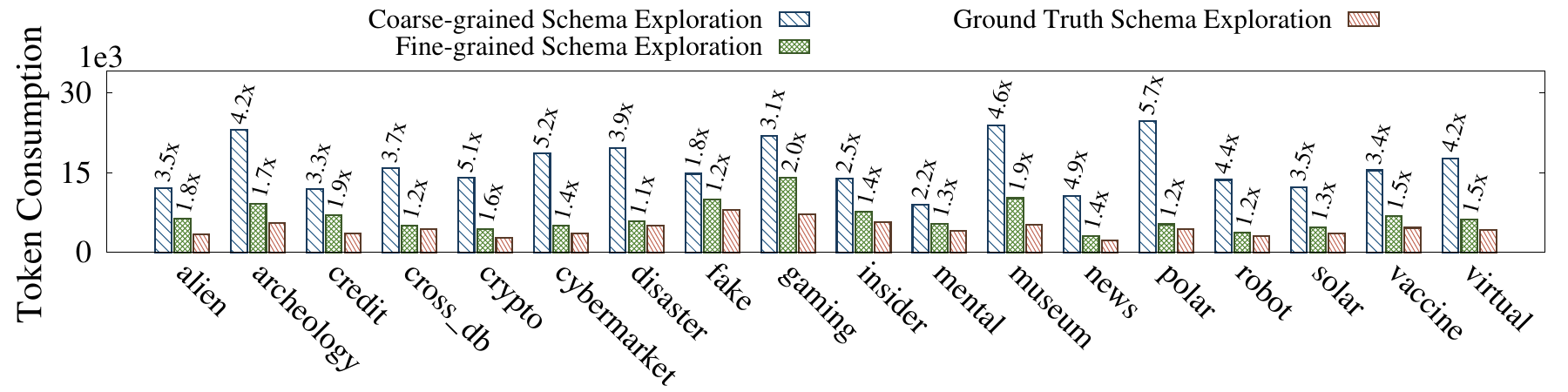}
        \captionof{figure}{Inefficiencies of coarse-grained exploration}
        \label{fig:token-ineff}
        \Description[Token consumption in coarse-grained schema exploration]{A plot showing the token inefficiencies of coarse-grained schema exploration.}
    \end{minipage}
\end{figure}

Based on our study, we observe a dichotomy between these choices. With coarse-grained APIs, the agent gets all the schema information, and has to then \emph{distill} down what is relevant. However, this leads to increased token cost (\S \ref{ineff-coarse-grained}) during exploration. Fine-grained APIs enable a more token-efficient exploration process, but the agent has no visibility into the full schema. Since the relevant tables are not known \emph{a priori}, constructing a correct query \emph{necessitates} exploring broadly, causing the agent to \emph{over-explore} and incorporate irrelevant tables into query construction, ultimately degrading accuracy. Furthermore, we summarize and show in Table \ref{tab:vendor-overview} that a majority of data systems expose fine-grained APIs, making any agent that is exposed to them prone to over-exploration and inaccurate SQL generation.

Our key insight for the effective use of fine-grained APIs is simple: let the environment provide the agent with \emph{directives} that help constrain its future exploration. Since over-exploration stems from exploring tables that later prove irrelevant, directives steer the agent away from such tables. Our initial results show that directives reduce over-exploration by up to 4.6$\times$ and boost SQL generation accuracy by up to 12.4\% ($\sim$4 percentage points). A key challenge is computing these directives and we present \sysname{}, a data system environment that addresses it.

\section{Motivation}
\label{motivation}

We now analyze the inefficiencies of coarse-grained API surfaces and quantify the problem of over-exploration.

\subsection{Setup}\label{setup}

Our data system environment is an MCP server for SQLite \cite{sqlite2020hipp} exposing the fine-grained API surface described in Table \ref{tab:api_surface}. We build our agent (\S \ref{agent-prompt}) using OpenCode \cite{anomaly_opencode_2025}, and evaluate it on 157 queries from the BIRD LiveSQLBench benchmark \cite{livesqlbench2025}. We evaluate with three models: GPT-5.4-mini, GPT-5.4, and Claude Sonnet 4.5, all configured using their default reasoning effort unless otherwise specified.

\subsection{Extent of Over-exploration}\label{over-exploration}

To measure over-exploration, we analyze the agent's tool call sequences. A tool call is \emph{exploratory} if it inspects a table's schema via \texttt{describe\_table} or a \texttt{read\_query} call of the form \texttt{PRAGMA table\_info (table\_name)}. We establish the ground truth exploration count by parsing each ground truth SQL query and extracting the number of unique tables it references, representing the exact number of schema explorations needed. Furthermore, to understand the effect a model's reasoning effort has on over-exploration, we measure it at each model's highest reasoning effort as well. 

\begin{figure}[H]
    \centering
    \includegraphics[width=0.8\linewidth]{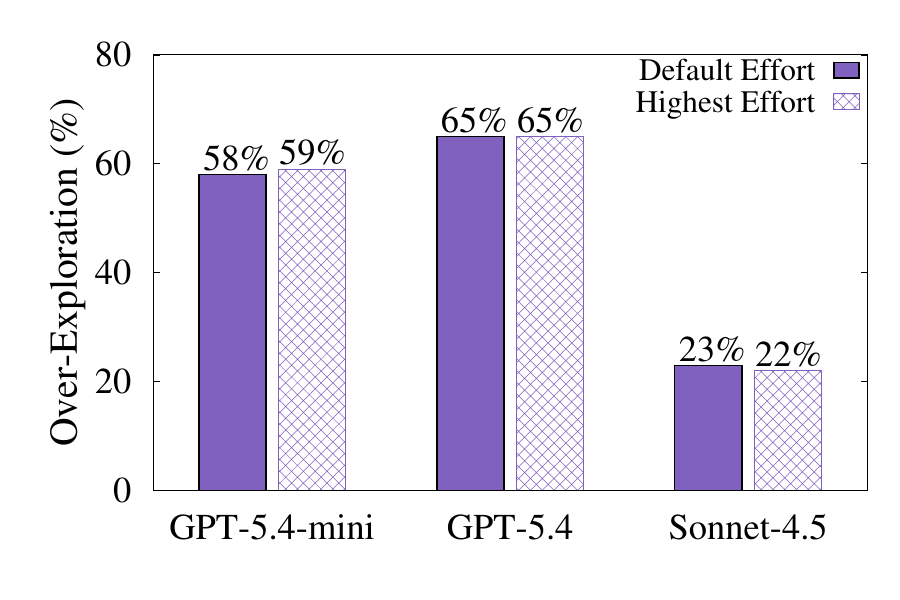}
    \caption{Over-exploration in data agents}
    \label{fig:over_explore}
    \Description[Over-exploration in data agents]{A plot comparing the extent of exploration versus ground truth in data agents.}
\end{figure}

Figure \ref{fig:over_explore} shows the percentage of exploratory calls across all queries above the ground truth. When exposed to fine-grained API surfaces, agents over-explore by up to 65\%, with reasoning effort having a negligible impact on over-exploration, suggesting that for each model, it is an inherent symptom when exposed to fine-grained API surfaces. More critically, over-exploration translates to inaccuracies in query construction as we discuss in \S \ref{effects}.

\subsection{Inefficiencies of Coarse-grained API Surfaces}\label{ineff-coarse-grained}

Since token consumption directly proxies cost, we measure the inefficiency of coarse-grained surfaces using Sonnet-4.5. With coarse-grained exploration, since the agent retrieves the entire schema for every query, we estimate its token consumption as the entire schema's token count times the number of queries. For comparison, we compute (1) the actual schema token consumption as the sum of schema tokens of tables used in the ground truth queries provided by BIRD, and (2) the total schema tokens of tables explored when the agent is exposed to a fine-grained API surface. Figure \ref{fig:token-ineff} compares these per database provided. Coarse-grained exploration consumes up to 5.7$\times$ more tokens than necessary while fine-grained exploration consumes up to 2$\times$ more, making coarse-grained exploration APIs highly cost-inefficient.

\begin{table}[t!]
    \centering
    \caption{API surface exposed by \sysname}
    \label{tab:api_surface}
    \small
    \renewcommand{\arraystretch}{1.05}
    \scalebox{0.9}{
    \begin{tabularx}{\linewidth}{>{\bfseries}l|X}
        API & \textbf{Description} \\
        \hline
        \texttt{list\_tables} & Return names of tables in the database \\
        \hline
        \texttt{describe\_table} & Return schema of a table \\
        \hline
        \texttt{get\_join\_info} & Return foreign key relationships between tables or for a specific table \\
        \hline
        \texttt{read\_query} & Execute read-only queries for data exploration and error validation \\
        \hline
        \texttt{submit\_query} & Submit the final SQL query \\
    \end{tabularx}}
\end{table}

\section{Design \& Implementation}
\label{design}

\begin{figure*}[!t]
    \centering
    \begin{subfigure}[t]{0.48\textwidth}
        \centering
        \includegraphics[width=\linewidth]{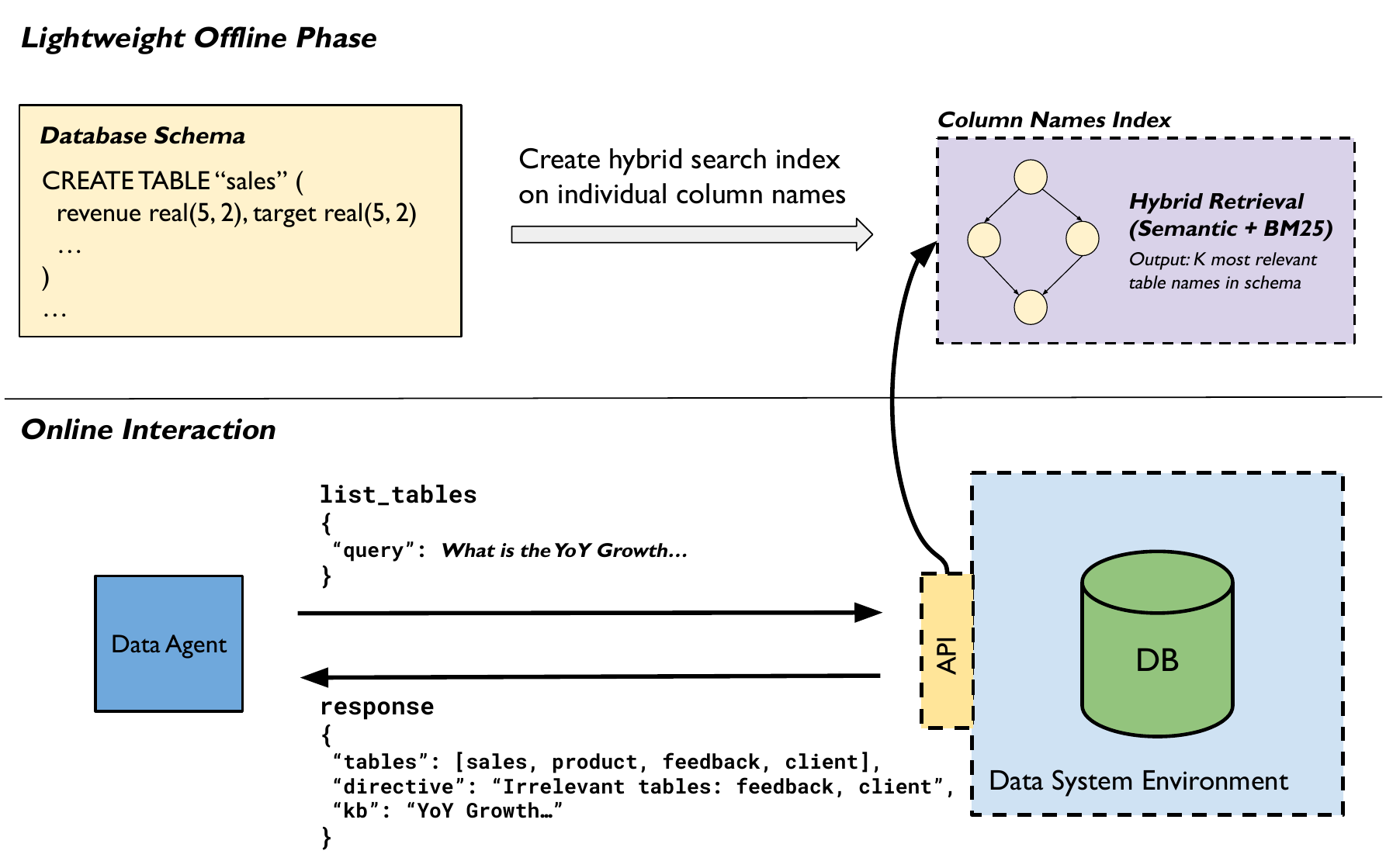}
        \caption{\sysname{} and the \emph{directive} mechanism}
        \label{fig:design-overview}
        \Description[Design of \sysname{}]{Design depicting the directive computation mechanism.}
    \end{subfigure}%
    \hfill
    \begin{subfigure}[t]{0.48\textwidth}
        \centering
        \includegraphics[width=\linewidth]{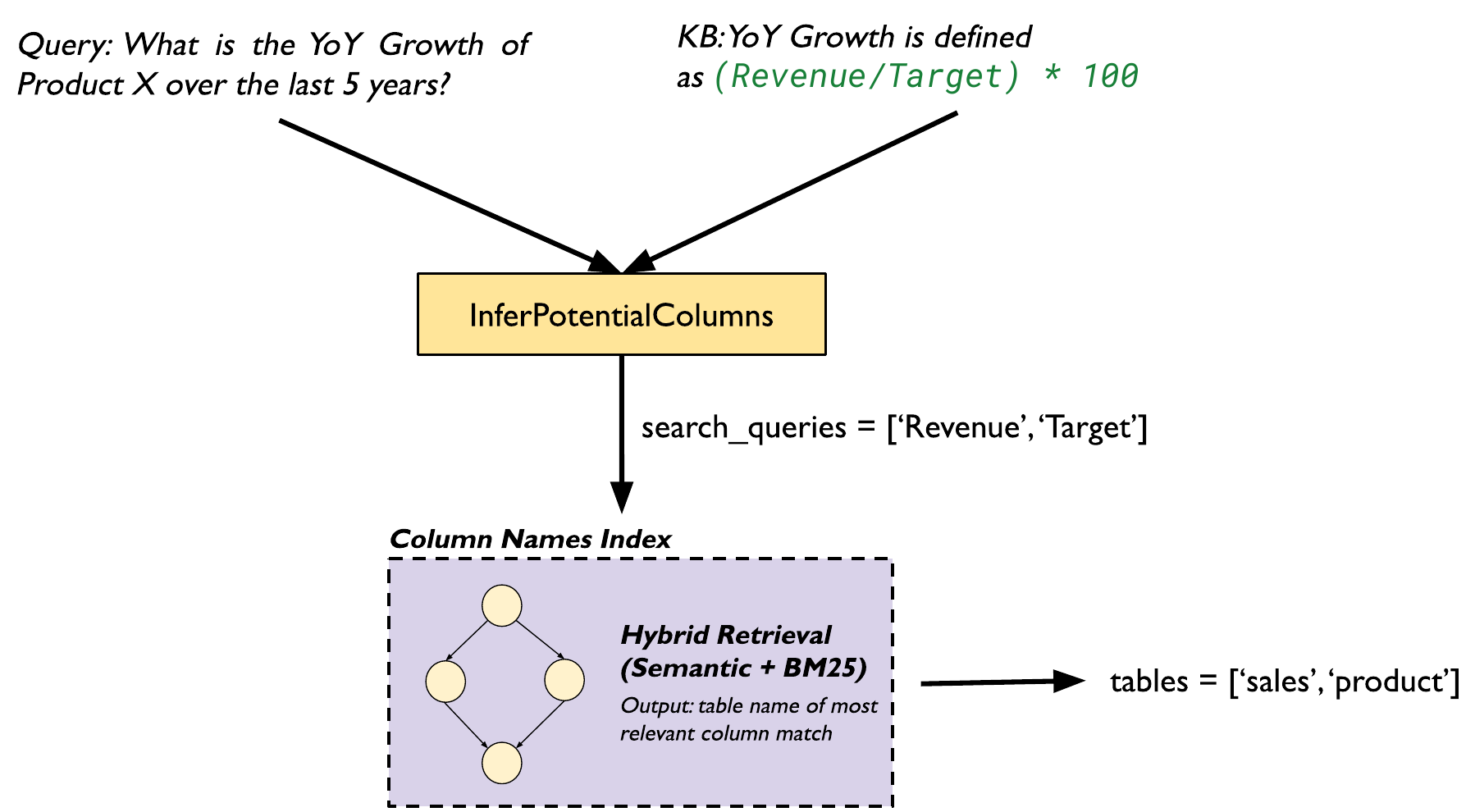}
        \caption{Computing directives from knowledge base entries}
        \label{fig:infer-tables}
        \Description[Computing directives]{Computing directives by inferring column names from knowledge base entries}
    \end{subfigure}
    \caption{Overview of \sysname{}'s design}
    \label{fig:design}
\end{figure*}

Over-exploration stems from exploring tables that prove to be irrelevant. Our key idea to address this issue is to compute and return {\em directives} that convey the set of potentially irrelevant tables as feedback to the agent to help guide its future exploration (as shown in Figure \ref{fig:design-overview}). The central challenge here is computing these directives. We design and implement \sysname{}, a data system environment that solves this.

Directives are returned with the response of \texttt{list\_tables}, since we observe this to consistently be the first tool call made by the agent. Interaction with \sysname{} requires a lightweight offline phase that constructs a hybrid search index (\hybrid{}) over individual column names, combining an IVF index for semantic similarity with a BM25 index \cite{robertson1994okapi} for keyword matches. The results are combined using the reciprocal rank fusion method \cite{RRF}. To generate embeddings for \hybrid{}, we use OpenAI's \texttt{text-embedding-3-large} model. As a result, \hybrid{} can be used to search for relevant column names and which table contains them. On searching \hybrid{}, it returns the table name of the most relevant match.

The key idea behind directives is that for the knowledge sources provided by the BIRD benchmark, directives are computed based on the observation that entries are defined in terms of relevant column names. Querying \hybrid{} using the entire knowledge base entry can lead to inaccuracies given multiple column names are likely to be present \cite{ma2023queryrewritingretrievalaugmentedlarge}. As a result, we first infer \emph{potential} column names by prompting (\S \ref{col-inf-prompt}) an inexpensive model as shown in Figure \ref{fig:infer-tables}. The inferred names now serve as search queries for \hybrid{}. Running retrieval for each of these gives us a set of potentially \emph{relevant} tables for this query. Directives are now the \emph{irrelevant} tables, computed as all table names that are \emph{not} part of this set.

\sysname{} is implemented as an MCP server following current data systems \cite{aws_aurora_dsql_mcp_2025, aws_mysql_mcp_2025, aws_postgres_mcp_2025, aws_redshift_mcp_2025, google_cloud_mcp_2025_bigquery, azure_dab_mcp_dml_2025, supabase_mcp_2025, neon_mcp_2026, motherduck_mcp_2025} and exposes the fine-grained API surface laid out in Table \ref{tab:api_surface}. We posit that directive computation should be tailored to the data system and the available context and that better mechanisms will further enhance the agent's success. \sysname{} balances offline and online costs and while schema linking approached \cite{10.1145/3725337,talaei2024chesscontextualharnessingefficient} make different cost tradeoffs, they can help achieve the same goal. Furthermore, companies building data agents \cite{openai2026dataagent, databricksgenie, metaagent} can leverage the semantically richer context available to them to compute directives more effectively.

\section{Evaluation}
\label{eval}

We use the setup described in \S \ref{setup}, with our agent overriding OpenCode's system prompts (\S \ref{agent-prompt}). We also disallow use of built-in tools thereby treating OpenCode as a simple agent loop with access to only the API surface exposed to it. Consequently, any agent implementation with MCP support \cite{anysphere_cursor_2024, anthropic_claude_code_2025, sourcegraph_amp_2025, anomaly_opencode_2025, openai_codex_cli_2025} can be used. Additionally, BIRD provides domain-specific knowledge bases to provide the agent with prerequisite context. Since retrieval of knowledge entries can be inaccurate \cite{ma2023queryrewritingretrievalaugmentedlarge}, we provide the agent with the exact entries needed for each query to isolate its effects and compare the following three systems on the basis of the directive mechanism: \emph{No Directives}, \sysname{}, \sysname-\emph{Oracle}.

\emph{No Directives}: This is a faithful representation of existing data system environments that expose fine-grained API surfaces. It exposes the API surface laid out in Table \ref{tab:api_surface}.

\sysname{}: This also exposes the API surface laid out in Table \ref{tab:api_surface}, computes directives and returns them as part of the response for \texttt{list\_tables} as described in \S \ref{design}. 

\sysname-\emph{Oracle}: This is identical to \sysname{} except it assumes an oracle present for computing fully accurate directives. The system looks at ground truth SQL queries to compute the set of irrelevant tables thereby guaranteeing 100\% accurate directives. With these, \sysname{}-\emph{Oracle} intends to represent the upper bound on agent performance when exposed to a fine-grained API surface.

\subsection{Over-exploration \& Its Effects}\label{effects}

\begin{figure}
    \centering
    \begin{subfigure}[c]{0.48\linewidth}
        \centering
        \includegraphics[width=\linewidth]{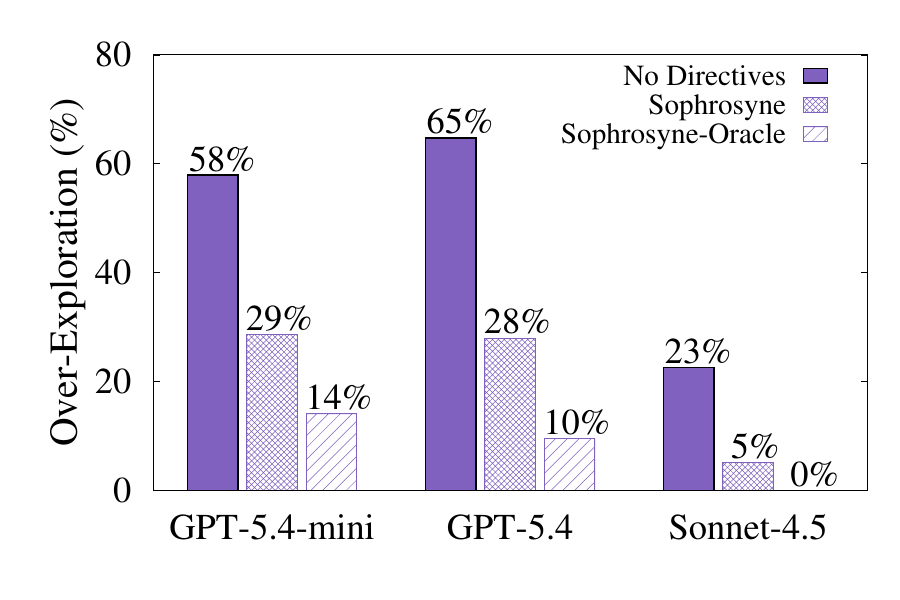}
        \caption{Measure of over-exploration}
        \label{fig:over_explore_variants}
        \Description[Over-exploration in data agents]{A plot comparing the extent of exploration versus ground truth in data agents across different systems.}
    \end{subfigure}
    \hfill
    \begin{subfigure}[c]{0.48\linewidth}
        \centering
        \includegraphics[width=\linewidth]{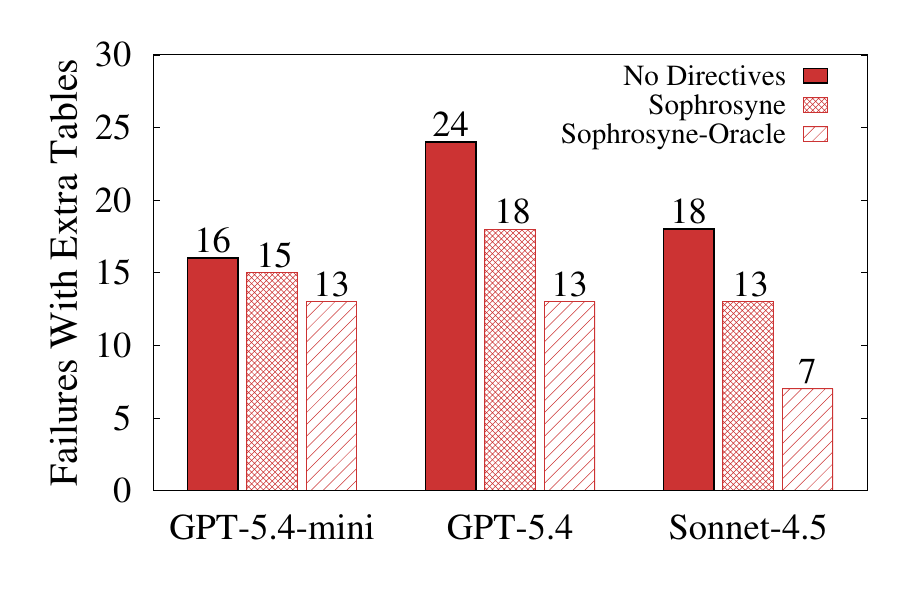}
        \caption{Failures with extra tables}
        \label{fig:failed_extra_variants}
        \Description[Failed queries with over-explored tables]{Comparing number of failed queries across systems containing over-explored tables}
    \end{subfigure}
    \caption{Quantifying over-exploration and its effects}
    \label{fig:over_explore_eval}
\end{figure}

We measure exploration using the methodology in \S \ref{over-exploration}. Figure \ref{fig:over_explore_variants} shows over-exploration as a percentage of exploratory calls above the required number. With directives, \sysname{} consistently reduces over-exploration across all models, and \sysname{}-\emph{Oracle} reduces it further, indicating that more accurate directive computation only yields better results.

Next, we examine how directives and over-exploration affect query accuracy. Figure \ref{fig:failed_extra_variants} shows the number of failed queries that reference tables absent from the ground truth. \emph{No Directives} serves as the baseline for these failures, though not all can be attributed to over-exploration, as logical errors in query formulation are still possible. With \sysname{}, these failures decrease consistently across all models. While we get improved accuracy with imperfect directives (\S \ref{directive-overhead}), \sysname-\emph{Oracle} further improves it with perfect directives indicating the effect over-exploration has on query accuracy.

For Sonnet-4.5, while \sysname{}-\emph{Oracle} eliminates over-exploration as seen in Figure \ref{fig:over_explore_variants}, we still see failures with extra tables (Figure \ref{fig:failed_extra_variants}) because although the agent may not explore these tables using explicit tool calls, it is still made aware of them via information returned by \texttt{get\_join\_info}.

\subsection{Execution Accuracy \& Token Cost}\label{exec-accuracy}

We use token counts as a proxy for inference cost \cite{openai2026pricing, anthropic2025pricing}. Table \ref{tab:accuracy_token_comparison} summarizes execution accuracy, input tokens and output tokens (including reasoning) across the three systems. \sysname{} consistently improves accuracy by reducing over-exploration, and since fewer schema elements are explored, the corresponding token counts also reduce. GPT-5.4-mini is an exception, incurring a 1.2\% output token overhead due to excess reasoning tokens and the model summarizing verbose SQL results before terminating for certain queries. We discuss cost in more detail in \S \ref{directive-overhead}. \sysname-\emph{Oracle} consistently improves execution accuracy and improves token cost, further indicating that better mechanisms to compute directives will only lead to better agent performance.

\begin{table}
  \centering
  \caption{Execution Accuracy and Token Cost}
  \medskip
  \label{tab:accuracy_token_comparison}
  \small
  \setlength{\tabcolsep}{3pt}
  \begin{tabular*}{\linewidth}{@{\extracolsep{\fill}}>{\bfseries}l|l|r|r|r}
    \textbf{Model} & \textbf{System} & \thead{\textbf{Exec.}\\ \textbf{Accuracy}} & \thead{\textbf{Input}\\ \textbf{Tokens}} & \thead{\textbf{Output}\\ \textbf{Tokens}} \\
      \hline
        \multirow{3}{*}{GPT-5.4-mini}
          & \emph{No Directives}        & 37.6\% & 7.67M & \grn{585K} \\
          & \sysname             & \grn{39.5\%} & \grn{7.60M} & 592K \\
          & \sysname-\emph{Oracle} & \darkgrn{41.4\%} & \darkgrn{6.67M} & \darkgrn{557K} \\
      \hline
      \multirow{3}{*}{GPT-5.4}
          & \emph{No Directives}          & 30.6\% & 6.28M & 406K \\
          & \sysname             & \grn{34.4\%} & \grn{6.15M} & \grn{404K} \\
          & \sysname-\emph{Oracle} & \darkgrn{37.6\%} & \darkgrn{5.04M} & \darkgrn{383K} \\
      \hline
      \multirow{3}{*}{Sonnet 4.5}
          & \emph{No Directives}          & 41.4\% & 8.74M & 343K \\
          & \sysname             & \grn{42.7\%} & \grn{8.00M} & \grn{318K} \\
          & \sysname-\emph{Oracle} & \darkgrn{45.9\%} & \darkgrn{6.72M} & \darkgrn{310K} \\
    \end{tabular*}
\end{table}

\subsection{Directive Accuracy \& System Overhead}\label{directive-overhead}

We now discuss the accuracy and overhead of our directive mechanism (\S \ref{design}). The offline phase embeds column names, costing \emph{one-tenth of a cent} totally. We treat this as negligible and ignore it in our discussion ahead. To measure accuracy of computed directives we calculate recall percentage with respect to the ground truth set of tables, and across runs of the agent, this averages to \textbf{82.5\%}. 

Table \ref{tab:cost_comparison} summarizes the base cost and overhead incurred with \sysname{}. The overhead incurred is \emph{8 cents}, covering cost to infer columns using GPT-5.4-nano and subsequently embed them (\S \ref{design}). \sysname{} is more cost-efficient except for GPT-5.4-mini, where the base cost reduction does not absorb the overhead. Sonnet-4.5 sees the largest gains, consistent with the over-exploration reduction in Figure \ref{fig:over_explore_variants}.

\begin{table}
  \centering
  \caption{Overhead of directive computation in \sysname}
  \medskip
  \label{tab:cost_comparison}
    \small
  \begin{tabularx}{\linewidth}{>{\bfseries}l|l|X|X|X}
    \textbf{Model} & \textbf{System} & \textbf{Base Cost} & \thead{\textbf{Over-}\\ \textbf{head}} & \textbf{Total} \\
      \hline
        \multirow{2}{*}{GPT-5.4-mini}
          & \emph{No Directives}   & \$8.39 & - & \grn{\$8.39} \\
          & \sysname & \$8.36 & \$0.08 & \$8.44 \\
      \hline
        \multirow{2}{*}{GPT-5.4}
          & \emph{No Directives}   & \$21.79 & - & \$21.79 \\
          & \sysname & \$21.43 & \$0.08 & \grn{\$21.51} \\
      \hline
        \multirow{2}{*}{Sonnet-4.5}
          & \emph{No Directives}   & \$31.36 & - & \$31.36 \\
          & \sysname & \$28.80 & \$0.08 & \grn{\$28.88} \\
    \end{tabularx}
\end{table}

\section{Related Work}
\label{related}

Text2SQL has a rich body of work ranging from building and fine-tuning custom models \cite{10.14778/3641204.3641221,10.1145/3654930,pourreza-rafiei-2024-dts,qin2025route,NGUYEN2025100135}, to using LLMs with sophisticated prompting strategies \cite{talaei2024chesscontextualharnessingefficient,shkapenyuk2025automaticmetadataextractiontexttosql} and inference time scaling \cite{deng2025reforcetexttosqlagentselfrefinement} to improve the accuracy of text2SQL. With LLMs becoming more adept at tool calling and interacting with data systems, works on agentic text2SQL focus on improving accuracy by synthesizing common failure modes in an offline phase and modifying the runtime workflow to provide relevant context to the agent \cite{biswal2026agentsmsemanticmemoryagentic,agarwal2026armingdataagentstribal}. Our work is the first to study the effect of the exposed API surface on text2SQL accuracy and identify over-exploration as a symptom of an overlooked tradeoff in API design.  \sysname{} addresses this by augmenting API responses, and can therefore be used with any existing agent and workflow, as well as in conjunction with other works, further improving accuracy.

\section{Future Work}
\label{sec-future}
We now outline directions for future work. Even with perfect directives, some models still over-explore, while those that avoid over-exploration can still produce failed queries with extra tables (Figure \ref{fig:over_explore_eval}) suggesting the effectiveness of directives depends on the model's instruction-following ability. Since directives are returned early, their influence may decay as the agent's context grows \cite{hong2025context} and returning \emph{reminders} when directives are not followed could help keep the model in line. Next, data agents answering queries spanning multiple data sources \cite{ma2026aiagentsanswerdata} need to discover these sources and subsequently relevant schema elements across them. With fine-grained surfaces, the effects of over-exploration would only be exacerbated. Directives aware of multiple sources would prove essential in this setting. Finally, in addition to guiding the exploration process, directives can also provide \emph{performance hints} to the agent in order to construct more efficient SQL. For example, nudging the agent to use a composite index's column prefix can dramatically improve query performance for large tables.

\clearpage
{
   \balance
   \bibliographystyle{plain}
   \bibliography{all-defs,ds,urls,all-confs}

@string{jan = "January" }

@string{mar = "March" }

@string{apr = "April" }

@string{may = "May" }

@string{jun = "June" }

@string{nov = "November" }

@inproceedings{lazylog,
  title="{LazyLog: A New Shared Log Abstraction for Low-Latency Applications}",
  author = {Xuhao Luo and Shreesha G. Bhat and Jiyu Hu and Ramnatthan Alagappan and Aishwarya Ganesan},
  crossref="SOSP24"
}

@misc{liu2025supportingaioverlordsredesigning,
      title={Supporting Our {AI} Overlords: Redesigning Data Systems to be Agent-First},
      author={Shu Liu and Soujanya Ponnapalli and Shreya Shankar and Sepanta Zeighami and Alan Zhu and Shubham Agarwal and Ruiqi Chen and Samion Suwito and Shuo Yuan and Ion Stoica and Matei Zaharia and Alvin Cheung and Natacha Crooks and Joseph E. Gonzalez and Aditya G. Parameswaran},
      year={2025},
      eprint={2509.00997},
      archivePrefix={arXiv},
      primaryClass={cs.AI},
      howpublished = {\url{https://arxiv.org/abs/2509.00997}},
}

@inproceedings{zaharia2025bridging,
  author = {Zaharia, Matei},
  title = {Bridging the Operational and Analytical Worlds with {Lakebase}},
  booktitle = {Proceedings of the 51st International Conference on Very Large Data Bases},
  series = {VLDB '25},
  year = {2025},
  location = {London, United Kingdom},
  note = {Keynote 3},
  url = {https://vldb.org/2025/files/keynote/vldb25-keynote3.pdf}
}

@misc{schick2023toolformerlanguagemodelsteach,
      title={{Toolformer}: Language Models Can Teach Themselves to Use Tools},
      author={Timo Schick and Jane Dwivedi-Yu and Roberto Dessì and Roberta Raileanu and Maria Lomeli and Luke Zettlemoyer and Nicola Cancedda and Thomas Scialom},
      year={2023},
      eprint={2302.04761},
      archivePrefix={arXiv},
      primaryClass={cs.CL},
      howpublished = {\url{https://arxiv.org/abs/2302.04761}},
}

@techreport{hong2025context,
  title = {Context Rot: How Increasing Input Tokens Impacts {LLM} Performance},
  author = {Hong, Kelly and Troynikov, Anton and Huber, Jeff},
  year = {2025},
  month = {July},
  institution = {Chroma},
  url = {https://research.trychroma.com/context-rot},
}

@misc{anthropic2025pricing,
  author = {Anthropic},
  title = {Pricing - {Claude} Documentation},
  year = {2025},
  howpublished = {\url{https://docs.claude.com/en/docs/about-claude/pricing}},
  note = {Accessed: 2025-12-11}
}

@misc{anthropic2024mcp,
  author = {Anthropic},
  title = {Introducing the {Model Context Protocol}},
  year = {2024},
  month = {November},
  howpublished = {\url{https://www.anthropic.com/news/model-context-protocol}},
  note = {Accessed: 2025-12-11}
}

@article{10.1145/3725337,
author = {Balaka, Muhammad Imam Luthfi and Alexander, David and Wang, Qiming and Gong, Yue and Krisnadhi, Adila and Castro Fernandez, Raul},
title = {{Pneuma}: Leveraging {LLMs} for Tabular Data Representation and Retrieval in an End-to-End System},
year = {2025},
issue_date = {June 2025},
publisher = {Association for Computing Machinery},
address = {New York, NY, USA},
volume = {3},
number = {3},
url = {https://doi.org/10.1145/3725337},
doi = {10.1145/3725337},
journal = {Proc. ACM Manag. Data},
month = jun,
articleno = {200},
numpages = {28}
}

@misc{shkapenyuk2025automaticmetadataextractiontexttosql,
      title={Automatic Metadata Extraction for Text-to-{SQL}},
      author={Vladislav Shkapenyuk and Divesh Srivastava and Theodore Johnson and Parisa Ghane},
      year={2025},
      eprint={2505.19988},
      archivePrefix={arXiv},
      primaryClass={cs.DB},
      howpublished = {\url{https://arxiv.org/abs/2505.19988}},
}

@misc{talaei2024chesscontextualharnessingefficient,
      title={{CHESS}: Contextual Harnessing for Efficient {SQL} Synthesis},
      author={Shayan Talaei and Mohammadreza Pourreza and Yu-Chen Chang and Azalia Mirhoseini and Amin Saberi},
      year={2024},
      eprint={2405.16755},
      archivePrefix={arXiv},
      primaryClass={cs.LG},
      howpublished = {\url{https://arxiv.org/abs/2405.16755}},
}

@misc{google_cloud_mcp_2025_bigquery,
  author       = {{Google Cloud}},
  title        = {Google Cloud {MCP} Servers: Big Query},
  year         = {2025},
  howpublished = {\url{https://docs.cloud.google.com/bigquery/docs/reference/mcp/tools_overview}},
  note         = {Preview. Accessed: 2026-03-15}
}

@misc{supabase_mcp_2025,
  author       = {{Supabase Community}},
  title        = {Supabase {MCP} Server},
  year         = {2025},
  publisher    = {GitHub},
  howpublished = {\url{https://github.com/supabase-community/supabase-mcp}},
  note         = {Apache-2.0 license. Accessed: 2026-03-15}
}

@misc{planetscale_mcp_2026,
  author       = {{PlanetScale}},
  title        = {{PlanetScale} {MCP} Server},
  year         = {2026},
  publisher    = {GitHub},
  howpublished = {\url{https://github.com/planetscale/mcp-server}},
  note         = {Apache-2.0 license. Accessed: 2026-03-15}
}

@misc{aws_aurora_dsql_mcp_2025,                                                                                  
author       = {{AWS Labs}},                                                                                   
title        = {{Aurora DSQL} {MCP} Server},                                                                   
year         = {2025},                                                                                         
publisher    = {AWS Labs},                                                                                      
howpublished = {\url{https://awslabs.github.io/mcp/servers/aurora-dsql-mcp-server}},
note         = {Apache-2.0 license. Accessed: 2026-04-19}
}

@misc{aws_mysql_mcp_2025,                                                                                        
author       = {{AWS Labs}},                            
title        = {{MySQL} {MCP} Server},                                                                         
year         = {2025},                                  
publisher    = {AWS Labs},
howpublished = {\url{https://awslabs.github.io/mcp/servers/mysql-mcp-server}},                                       
note         = {Apache-2.0 license. Accessed: 2026-04-19}
}

@misc{aws_postgres_mcp_2025,                                                                                     
author       = {{AWS Labs}},                            
title        = {{Postgres} {MCP} Server},
year         = {2025},
publisher    = {AWS Labs},                                                                                       
howpublished = {\url{https://awslabs.github.io/mcp/servers/postgres-mcp-server}},
note         = {Apache-2.0 license. Accessed: 2026-04-19}                                                      
}

@misc{aws_redshift_mcp_2025,                                                                                     
author       = {{AWS Labs}},                            
title        = {{Redshift} {MCP} Server},
year         = {2025},                                                                                         
publisher    = {AWS Labs},
howpublished = {\url{https://awslabs.github.io/mcp/servers/redshift-mcp-server}},                                    
note         = {Apache-2.0 license. Accessed: 2026-04-19}                                                      
}

@misc{azure_dab_mcp_dml_2025,
    author       = {{Microsoft}},
    title        = {{Azure SQL MCP Tools}},
    year         = {2025},
    publisher    = {Microsoft Learn},
    howpublished = {\url{https://learn.microsoft.com/en-us/azure/data-api-builder/mcp/data-manipulation-language-tools}},
    note         = {Accessed: 2026-04-19}
}

@misc{motherduck_mcp_2025,
    author       = {{MotherDuck}},
    title        = {{MotherDuck}'s {DuckDB} {MCP} Server},
    year         = {2025},
    publisher    = {GitHub},
    howpublished = {\url{https://github.com/motherduckdb/mcp-server-motherduck}},
    note         = {MIT license. Accessed: 2026-04-19}
}

@misc{google_spanner_mcp_2026,
  author       = {{Google Cloud}},
  title        = {Use the {Spanner} Remote {MCP} Server},
  year         = {2026},
  howpublished = {\url{https://docs.cloud.google.com/spanner/docs/use-spanner-mcp}},
  note         = {Accessed: 2026-03-15}
}

@misc{turso_mcp_2026,
  author       = {{Turso}},
  title        = {Introducing the {Turso} Database {MCP}},
  year         = {2025},
  howpublished = {\url{https://turso.tech/blog/introducing-the-turso-database-mcp-server}},
  note         = {Accessed: 2026-05-03}
}

@misc{anthropic_claude_code_2025,
  author       = {{Anthropic}},
  title        = {Claude Code},
  year         = {2025},
  howpublished = {\url{https://code.claude.com/docs/en/overview}},
  note         = {Accessed: 2026-03-15}
}

@misc{anysphere_cursor_2024,
  author       = {{Anysphere}},
  title        = {Cursor: The {AI} Code Editor},
  year         = {2024},
  howpublished = {\url{https://cursor.com}},
  note         = {Accessed: 2026-03-15}
}

@misc{openai_codex_cli_2025,
  author       = {{OpenAI}},
  title        = {Codex {CLI}},
  year         = {2025},
  howpublished = {\url{https://developers.openai.com/codex/cli/}},
  note         = {Accessed: 2026-03-15}
}

@misc{anomaly_opencode_2025,
  author       = {{Anomaly (SST)}},
  title        = {{OpenCode}: The Open Source {AI} Coding Agent},
  year         = {2025},
  howpublished = {\url{https://opencode.ai}},
  note         = {Apache-2.0 license. Accessed: 2026-03-15}
}

@misc{sourcegraph_amp_2025,
  author       = {{Sourcegraph}},
  title        = {Amp: {AI} Coding Agent},
  year         = {2025},
  howpublished = {\url{https://ampcode.com}},
  note         = {
                  Accessed: 2026-03-15}
}

@misc{livesqlbench2025,
  author       = {BIRD Team},
  title        = {{LiveSQLBench}: A Dynamic and Contamination-Free Benchmark for Evaluating {LLMs} on Real-World Text-to-{SQL} Tasks},
  year         = {2025},
  howpublished = {\url{https://github.com/bird-bench/livesqlbench}},
  note         = {Accessed: 2025-05-22}
}

@misc{sqlite2020hipp,
  title={{SQLite}},
  howpublished={\url{https://www.sqlite.org/index.html}},
  note={Version 3.31.1},
  year={2020},
  author={Hipp, Richard D.}
}

@misc{biswal2026agentsmsemanticmemoryagentic,
      title={{AgentSM}: Semantic Memory for Agentic Text-to-{SQL}},
      author={Asim Biswal and Chuan Lei and Xiao Qin and Aodong Li and Balakrishnan Narayanaswamy and Tim Kraska},
      year={2026},
      eprint={2601.15709},
      archivePrefix={arXiv},
      primaryClass={cs.AI},
      howpublished = {\url{https://arxiv.org/abs/2601.15709}},
}

@misc{agarwal2026armingdataagentstribal,
      title={Arming Data Agents with Tribal Knowledge},
      author={Shubham Agarwal and Asim Biswal and Sepanta Zeighami and Alvin Cheung and Joseph Gonzalez and Aditya G. Parameswaran},
      year={2026},
      eprint={2602.13521},
      archivePrefix={arXiv},
      primaryClass={cs.DB},
      howpublished = {\url{https://arxiv.org/abs/2602.13521}},
}

@misc{openai2026dataagent,
  title   = {Inside Our In-House Data Agent},
  author  = {{OpenAI}},
  year    = {2026},
  month   = jan,
  day     = {29},
  howpublished = {\url{https://openai.com/index/inside-our-in-house-data-agent/}},
}

@inproceedings{robertson1994okapi,
  title     = {Some Simple Effective Approximations to the 2-Poisson Model for Probabilistic Weighted Retrieval},
  author    = {Robertson, Stephen E. and Walker, Steve},
  booktitle = {Proceedings of the 17th Annual International ACM SIGIR Conference on Research and Development in Information Retrieval},
  pages     = {232--241},
  year      = {1994},
  publisher = {Springer-Verlag},
  address   = {Dublin, Ireland},
}

@misc{openai2026pricing,
  title        = {{OpenAI API} Pricing},
  author       = {{OpenAI}},
  year         = {2026},
  howpublished = {\url{https://developers.openai.com/api/docs/pricing}},
  note         = {Accessed: 2026-04-02},
}

@misc{ma2023queryrewritingretrievalaugmentedlarge,
      title={Query Rewriting for Retrieval-Augmented Large Language Models},
      author={Xinbei Ma and Yeyun Gong and Pengcheng He and Hai Zhao and Nan Duan},
      year={2023},
      eprint={2305.14283},
      archivePrefix={arXiv},
      primaryClass={cs.CL},
      howpublished = {\url{https://arxiv.org/abs/2305.14283}},
}

@misc{deng2025reforcetexttosqlagentselfrefinement,
      title={{ReFoRCE}: A Text-to-{SQL} Agent with Self-Refinement, Consensus Enforcement, and Column Exploration},
      author={Minghang Deng and Ashwin Ramachandran and Canwen Xu and Lanxiang Hu and Zhewei Yao and Anupam Datta and Hao Zhang},
      year={2025},
      eprint={2502.00675},
      archivePrefix={arXiv},
      primaryClass={cs.CL},
      howpublished = {\url{https://arxiv.org/abs/2502.00675}},
}

@misc{neon_mcp_2026,
  author       = {{Neon Database}},
  title        = {MCP server for interacting with Neon Management API and databases},
  year         = {2026},
  publisher    = {GitHub},
  howpublished = {\url{https://github.com/neondatabase/mcp-server-neon}},
  note         = {MIT license. Accessed: 2026-03-15}
}

@inproceedings{RRF,
author = {Cormack, Gordon V. and Clarke, Charles L A and Buettcher, Stefan},
title = {Reciprocal rank fusion outperforms condorcet and individual rank learning methods},
year = {2009},
isbn = {9781605584836},
publisher = {Association for Computing Machinery},
address = {New York, NY, USA},
url = {https://doi.org/10.1145/1571941.1572114},
doi = {10.1145/1571941.1572114},
booktitle = {Proceedings of the 32nd International ACM SIGIR Conference on Research and Development in Information Retrieval},
pages = {758–759},
numpages = {2},
location = {Boston, MA, USA},
series = {SIGIR '09}
}

@misc{databricksgenie,
  title   = {Introducing Genie Agent Mode},
  author  = {{Databricks}},
  year    = {2026},
  month   = apr,
  day     = {17},
  howpublished = {\url{https://www.databricks.com/blog/introducing-genie-agent-mode}},
}

@misc{metaagent,
  title   = {Inside Meta’s Home Grown AI Analytics Agent},
  author  = {{Analytics at Meta}},
  year    = {2026},
  month   = mar,
  day     = {30},
  howpublished = {\url{https://medium.com/@AnalyticsAtMeta/inside-metas-home-grown-ai-analytics-agent-4ea6779acfb3}},
}

@misc{mcpcli,
  title   = {Building agents that reach production systems with MCP},
  author  = {{Claude Platform}},
  year    = {2026},
  month   = apr,
  day     = {22},
  howpublished = {\url{https://claude.com/blog/building-agents-that-reach-production-systems-with-mcp}},
}

@misc{mcpauth,
  title   = {Building agents that reach production systems with MCP},
  author  = {{Model Context Protocol}},
  howpublished = {\url{https://modelcontextprotocol.io/docs/tutorials/security/authorization}},
}

@misc{ma2026aiagentsanswerdata,
      title={Can AI Agents Answer Your Data Questions? A Benchmark for Data Agents}, 
      author={Ruiying Ma and Shreya Shankar and Ruiqi Chen and Yiming Lin and Sepanta Zeighami and Rajoshi Ghosh and Abhinav Gupta and Anushrut Gupta and Tanmai Gopal and Aditya G. Parameswaran},
      year={2026},
      eprint={2603.20576},
      archivePrefix={arXiv},
      primaryClass={cs.DB},
      url={https://arxiv.org/abs/2603.20576}, 
}

@article{10.14778/3641204.3641221,
author = {Gao, Dawei and Wang, Haibin and Li, Yaliang and Sun, Xiuyu and Qian, Yichen and Ding, Bolin and Zhou, Jingren},
title = {Text-to-SQL Empowered by Large Language Models: A Benchmark Evaluation},
year = {2024},
issue_date = {January 2024},
publisher = {VLDB Endowment},
volume = {17},
number = {5},
issn = {2150-8097},
url = {https://doi.org/10.14778/3641204.3641221},
doi = {10.14778/3641204.3641221},
journal = {Proc. VLDB Endow.},
month = jan,
pages = {1132–1145},
numpages = {14}
}

@article{10.1145/3654930,
author = {Li, Haoyang and Zhang, Jing and Liu, Hanbing and Fan, Ju and Zhang, Xiaokang and Zhu, Jun and Wei, Renjie and Pan, Hongyan and Li, Cuiping and Chen, Hong},
title = {CodeS: Towards Building Open-source Language Models for Text-to-SQL},
year = {2024},
issue_date = {June 2024},
publisher = {Association for Computing Machinery},
address = {New York, NY, USA},
volume = {2},
number = {3},
url = {https://doi.org/10.1145/3654930},
doi = {10.1145/3654930},
journal = {Proc. ACM Manag. Data},
month = may,
articleno = {127},
numpages = {28}
}

@inproceedings{pourreza-rafiei-2024-dts,
    title = "{DTS}-{SQL}: Decomposed Text-to-{SQL} with Small Large Language Models",
    author = "Pourreza, Mohammadreza  and
      Rafiei, Davood",
    editor = "Al-Onaizan, Yaser  and
      Bansal, Mohit  and
      Chen, Yun-Nung",
    booktitle = "Findings of the Association for Computational Linguistics: EMNLP 2024",
    month = nov,
    year = "2024",
    address = "Miami, Florida, USA",
    publisher = "Association for Computational Linguistics",
    url = "https://aclanthology.org/2024.findings-emnlp.481/",
    doi = "10.18653/v1/2024.findings-emnlp.481",
    pages = "8212--8220"
}

@inproceedings{qin2025route,
title={{ROUTE}: Robust Multitask Tuning and Collaboration for Text-to-{SQL}},
author={Yang Qin and Chao Chen and Zhihang Fu and Ze Chen and Dezhong Peng and Peng Hu and Jieping Ye},
booktitle={The Thirteenth International Conference on Learning Representations},
year={2025},
url={https://openreview.net/forum?id=BAglD6NGy0}
}

@article{NGUYEN2025100135,
title = {Fine-tuning text-to-SQL models with reinforcement-learning training objectives},
journal = {Natural Language Processing Journal},
volume = {10},
pages = {100135},
year = {2025},
issn = {2949-7191},
doi = {https://doi.org/10.1016/j.nlp.2025.100135},
url = {https://www.sciencedirect.com/science/article/pii/S2949719125000111},
author = {Xuan-Bang Nguyen and Xuan-Hieu Phan and Massimo Piccardi}
}
}

\clearpage
\nobalance
\appendix
\section{System Prompts}
\label{appendix}

\subsection{Column Inference System Prompt}\label{col-inf-prompt}

Below is the system prompt used for column inference in the directive computing mechanism (\S \ref{design}).

\begin{tcolorbox}[
  colback=gray!5,
  colframe=gray!75,
  title=Column Inference System Prompt,
  fonttitle=\bfseries,
  breakable,
  before skip=4pt,
  lines before break=3
]
\small
You extract column-search terms for schema/table filtering.

\vspace{0.5em}
\textbf{Task:} Given a natural-language database question and optional knowledge-base context for one fact or formula, return a JSON array of short phrases that are most likely to match column names, column descriptions, or column meanings. These phrases will be used to retrieve relevant columns and tables.

\vspace{0.5em}
\textbf{Primary objective:} Maximize recall of relevant column concepts without adding generic SQL noise.

\vspace{0.5em}
\textbf{Include:}
\begin{itemize}[leftmargin=1.5em, itemsep=0pt, topsep=2pt]
  \item Domain metrics, measures, statuses, categories, thresholds, units, dimensions, or attributes needed to answer the question
  \item Concepts used for filtering, grouping, ordering, joining, or computing the answer
  \item Variables or operands explicitly named in the KB context if they would need backing columns
  \item Specific qualified concepts rather than broad parents
  \item A likely schema-style variant only when it is materially different and useful
\end{itemize}

\vspace{0.5em}
\textbf{Do not include:}
\begin{itemize}[leftmargin=1.5em, itemsep=0pt, topsep=2pt]
  \item SQL operation words such as \texttt{count}, \texttt{sum}, \texttt{average}, \texttt{max}, \texttt{min}, \texttt{top}, \texttt{sort}, \texttt{group}
  \item Generic verbs or filler words such as show, list, find, calculate, value, record, row
  \item Table names or entity names unless they are also plausible column concepts
  \item Duplicated broad + specific pairs; keep the more specific phrase
  \item Explanations, labels, or any text outside the JSON array
\end{itemize}

\vspace{0.5em}
\textbf{Normalization rules:}
\begin{itemize}[leftmargin=1.5em, itemsep=0pt, topsep=2pt]
  \item Output 0 to 8 items
  \item Order from most important to least important
  \item Use short lowercase noun phrases
  \item Prefer singular forms when natural
  \item No punctuation except meaningful abbreviations
  \item Return only raw JSON
\end{itemize}

\vspace{0.5em}
\textbf{Examples}

\vspace{0.3em}
\textit{Question:} Which observations have approximate air quality above safe levels? \\
\textit{Output:} \texttt{["approximate air quality", "safe level"]}

\vspace{0.3em}
\textit{Question:} Find the largest circumference. \\
\textit{Context:} circumference = 2 $\times$ CircleConstant $\times$ Radius / UnitOfMeasure \\
\textit{Output:} \texttt{["circumference", "circle constant", "radius", "unit of measure"]}

\vspace{0.3em}
\textit{Question:} Show facilities with severe corrosion risk. \\
\textit{Output:} \texttt{["severe corrosion risk"]}

\vspace{0.3em}
Return only the JSON array.
\end{tcolorbox}

\subsection{Agent System Prompt}\label{agent-prompt}

Below is the system prompt using which the OpenCode agent is constructed.

\begin{tcolorbox}[
  colback=gray!5,
  colframe=gray!75,
  title=Agent System Prompt,
  fonttitle=\bfseries,
  breakable,
  before skip=4pt,
  lines before break=3
]
\small
You are a confident data analyst assistant with access to a SQLite database.
Your task is to answer the user's question by querying the database.

\vspace{0.5em}
\textbf{Available tools:}
\begin{itemize}[leftmargin=1.5em, itemsep=0pt, topsep=2pt]
  \item \texttt{list\_tables}: List all tables in the database.
  \item \texttt{describe\_table}: Get schema information for a single table.
    \item \texttt{get\_join\_info}: Get all foreign key relationships between tables to understand how tables relate to each other.
  \item \texttt{read\_query}: Execute a SELECT query for exploration and intermediate checks.
  \item \texttt{submit\_final\_query}: Submit your final SQL query that answers the user's question. Only call this when you are confident in your answer.
\end{itemize}

\vspace{0.5em}
\textbf{Strategy:}
\begin{enumerate}[leftmargin=1.5em, itemsep=0pt, topsep=2pt]
  \item Execute appropriate queries to find the information needed.
  \item Make sure the queries you issue don't return massive intermediate outputs since that might exhaust the context window.
  \item Provide a clear, concise answer to the user's query, provide only what is asked. No explanations. No insights.
\end{enumerate}

\vspace{0.5em}
\textbf{Critical:}
\begin{enumerate}[leftmargin=1.5em, itemsep=0pt, topsep=2pt]
  \item Use \texttt{max\_rows} whenever possible with \texttt{read\_query} to minimise the amount returned by the server.
  \item Use ONLY SQL queries to arrive at the final result. All formatting should happen via SQL. No processing.
  \item You MUST use \texttt{submit\_final\_query} and ONLY for your final answer query.
  \item Once you are instructed to terminate, do so immediately.
\end{enumerate}

\vspace{0.3em}
Be methodical in your exploration.
\end{tcolorbox}

\end{document}